\documentstyle[12pt]{article}
\begin{document}
%
%
\title{Unified treatment of $A+B\rightarrow 0\ $ and $A+A\rightarrow
0$ chemical reactions through Thompson's approach.}
\author{Cl\'{a}udio Nassif~\footnote{e-mail: cnassif@fisica.ufmg.br} and  P.R.Silva}
\maketitle
\begin{center}
Departamento de F\'{\i}sica - ICEx - UFMG \\
Caixa Postal 702 - 30.123-970 - Belo Horizonte - MG - Brazil
\end{center}
\begin{abstract}
\footnotesize{
In this work we propose an action to describe diffusion
limited chemical reactions belonging to various classes of universality.
This action is treated through Thompson's approach and can encompass both
cases where we have segregation as in the $A+B\rightarrow 0$
reaction, as well the simplest one, namely the $A+A\rightarrow 0$ reaction
.Our results for long time and long wavelength behaviors of the species
concentrations and reaction rates agree with exact results of Peliti for
$A+A\rightarrow 0$ reaction and rigorous results of Bramson and Lebowitz for
$A+B\rightarrow 0$ reaction, with equal initial concentrations.The
different classes of universality are reflected by the obtained upper
critical dimensions varying continuously from $d_{c}=2$ in the first case to 
$d_{c}=4$ in the last one. Just at the upper critical dimensions we find
universal logarithmic corrections to the mean field behavior.}
\end{abstract}

\section{Introduction}

Since the pioneering work of Ovchinnikov and Zeldovich (O Z)\cite{1},
followed by Toussaint and Wilczek \cite{2}, the study of the chemical
reactions between two diffusing species which annihilate each other or
combine to form an inert product when they meet, has becomming a subject of
increasing interest [1 to 8]. As has been recognized since O.
Z. work\cite{1} (see also [2] and [3]), the spatial fluctuations in the
particle density plays a basic role in the long time behavior of their
concentrations. Then classical kinetics of reactions rates become the proper
description of this matter, only above some upper critical dimension.

Early treatments considered situations starting with equal or unequal
homogeneous concentrations of the two reactant species [1 to 8], but more
recent studies have also been adressed to the case of inhomogeneous initial
conditions [9 to 11].

A simple example of diffusion limited reaction is given by $A+A\rightarrow \
0$ (inert product). A field theory description of this reaction was
proposed by Peliti [12], performing the renormalization of it at all orders
in the perturbation theory.

Krug [13] has proposed a continuous version of a diffusion limited
annihilation (DLA) reaction with a point source. Based on the work of Krug
[13], one of the present authors [14] wrote na action that when treated by
the method of the dimensions due to Thompson [15], reproduced the exact
results of Peliti [12] for the $A+A\rightarrow ~ 0$ reaction. Thompson's
method [15] could be thougth as an alternative approach to the
renormalization group formalism [16].

Some important results obtained by Peliti [12] is that the long time behavior
of the concentration of species A is given by

\begin{equation}
\langle\epsilon \rangle \sim \left\{
\begin{array}{ll}
t^{-\frac{d}{2}},& d < 2 \\ 
\frac{\ln (t)}{t},& d = 2 \\ 
t^{-1},& d > 2,
\end{array}
\right.
\end{equation} 
where $d_{c}=2$ is the upper critical dimension, namely the dimension above
which the system behaves classically.

\bigskip

On the other hand Bramson and Lebowitz[5] performed a rigorous treatment
looking for the long time behavior of the concentrations $\langle
\epsilon _{A}\rangle $ and $\langle \epsilon _{B}\rangle $ of two different
species in the reaction $A+B\rightarrow 0$, both for equal and unequal
initial concentrations of the two species. In the case of equal initial
concentrations, namely when $\langle $ $\epsilon _{A}\rangle _{0}
=\langle \epsilon _{B}\rangle _{0}$, they found that:

\begin{equation}
\langle \epsilon_A \rangle = \langle \epsilon_B \rangle \sim \left\{
\begin{array}{ll}
t^{-\frac{d}{4}}, & d < 4 \\ 
t^{-1}, & d > 4,
\end{array}
\right.
\end{equation}
being $d_{c}=4$ the upper critical dimension in this kind of two species
annihilation reaction.

A heuristic method as a means to obtain scaling laws for $A+A\rightarrow ~ 0$
and $A+B\rightarrow ~ 0$ reactions was developed by Lindenberg, Shen and
Kopelman (LSK [17]). LSK work was concerned with the study of the above
reactions both in euclidian and in fractal geometries and they found that

\begin{equation}
\langle \epsilon \rangle \sim t^{-\alpha}
\end{equation}

where

\begin{equation}
\alpha = \frac{d_{s}}{2} \left[ 1-\frac{\gamma d_{s}}{2d_{f}}\right]
\end{equation}

In (4), $d_{f}$ and $d_{s}$ are respectively fractal and spectral dimensions,
and the proper choise of $\gamma$ leads to the description of a specific kind of
reaction. As example, in euclidian geometries, putting $d_{s}=d_{f}=d$,
we must choose $\gamma =1$ for the $A+B\rightarrow ~ 0$ reaction and
$ \gamma =0$ for the $A+A\rightarrow ~ 0$ reaction.

The aim of this paper is to apply Thompson`s method [15] to study the long
time behavior of some different kinds of diffusion limited reactions.

C. J. Thompson [15] proposed a simple heuristic method as a means to study
the critical behavior of a system undergoing second order phase transition.
He started from a Landau-Ginsburg-Wilson free energy or hamiltonian, and
was able to get an explicit relation for the correlation length critical
exponent ($\nu$) as a function of the lattice dimensionality $(d)$. If we think
that this $\Phi ^{4}$- theory is within the same class of universality of
the Ising model, Thompson`s work reproduces the exact results for $\nu
(d=2) $ $=1$ and $\nu (d=1)\longrightarrow \infty $.

Thompson`s method has been applied to obtain the correlation length critical
exponent of the Random Field Ising Model by Aharony, Imry and Ma [18] and
by one of the present authors [19]. His method was also used to evaluate
the correlation critical exponent of the N-vector Model [20]. Yang-Lee
Edge Singularity Critical Exponents [21] has been also studied by this
method.
In section 2, we propose an action to describe the $A+B\rightarrow 0$
reaction and we treat it through the Thompson 's approach. This action has
been evaluated as an extension of that employed in a previous work [14] in
order to treat the $A$ $+A\rightarrow 0$ reaction, and in doing this, an
important novel ingredient to be considered is segregation. As we will see
our treatment is able to reproduce some of Bramson and Lebowitz results [5].
In section 3 we pursue further on this subject, and propose a generalized
action which reproduces the achievements of section 2 and those of the
previous work [14] as particular cases. We also will show that this
generalization could encompass more general situations as some of those
described in LSK [17] paper.
Finally we conclude in section 4 with some comments and prospects.

\section{The $\boldmath A+B \rightarrow 0$ Reaction.}

In this section we propose to apply Thompson's approach as a means to study
the two species annihilation reaction $(A+B$ $\rightarrow 0)$, with equal
initial concentrations.

For the case of two species annihilation, at first sight, we would need
two differential equations to describe its chemical kinetics. However, for
equal initial densities, due to the simmetry of the problem, the two
equations merge into a single one. Of course, if the concentrations are
equals at starting, they remain this way during the passage of the time.

The starting point of this work is the differential equation:

\begin{equation}
\frac{\partial \epsilon \left( r,t\right) }{\partial t}=D \nabla
^{2}\epsilon +h\langle \epsilon \rangle -K\epsilon ^{2},
\end{equation}
where $\epsilon $ is the concentration of species(A or B), $D$ is the
diffusion constant, $h \langle \epsilon \rangle$ stands for an effective
input rate and $K$ is the reaction rate. The novelty presented by equation
(5) as compared to the continuous approximation of the $A$ $+A$ $\rightarrow
0$ reaction introduced by Krug [13], is the presence of the effective input
rate $h$ $\langle \epsilon \rangle $. This term has been introduced in the
present work as a way to account for the phenomenon of segregation, because
we know that the mean value $\langle \epsilon \rangle $ decays with the
time in such a way that the effective input rate $h \langle \epsilon
\rangle $ also decreases in time.

As it is well known, segregation appears in these reactions due to the
impossibility of self-annihilation between A (or B) species.

\bigskip

In order to treat the phenomena described by (5) using Thompson's approach,
let us define the action:

\begin{equation}
A=\int_{l^{d}}d^{d}r\left[ \frac{1}{2}D\left( \nabla \epsilon \right)
^{2}-h\langle \epsilon \rangle \epsilon +\frac{1}{3}K\epsilon ^{3}+\frac{1}{2
}\frac{\partial \left( \epsilon ^{2}\right) }{\partial t}\right].
\end{equation}
Equation (5) can be obtained from (6), by imposing $\delta A=0$, for
constant $t$.

In an analogous way to the Thompson's reasoning [15] we make the following
heuristic assumptions:

\begin{description}
\item[a-] When the integral in (6) is taken over the cube $l^{d}$ in $d$
dimensions, the absolute values of the four terms separately in (6) are all
of order unity.
\item[b-] $K(l)$ is finite in the limit $l\rightarrow \infty $.
\end{description}

Using assumption (a),that is a scaling assumption, in the first term of (6)
we have:
\begin{equation}
\left| \int_{l^{d}}d^{d}r\left[ \frac{1}{2}D\left( \nabla \epsilon \right)
^{2}\right] \right| \sim l^{d-2}\langle \epsilon ^{2}\rangle \sim 1,
\end{equation}
so that the mean squared value of $\epsilon $ ($\langle \epsilon
^{2}\rangle $) behaves as
\begin{equation}
\langle \epsilon ^{2}\rangle \sim l^{2-d}.
\end{equation}
For the fourth term in (6),we have
\begin{equation}
\left| \int_{l^{d}}d^{d}r\left[ \frac{1}{2}\frac{\partial \left( \epsilon
^{2}\right) }{\partial t}\right] \right| \sim \langle \Omega \rangle
\langle \epsilon ^{2}\rangle l^{d}\sim 1
\end{equation}
where we have considered that
\begin{equation}
\epsilon \left( t\right) =\epsilon _{0}\exp \left( -\Omega t\right)
\end{equation}

Putting (8) into (9), we obtain:
\begin{equation}
\langle \Omega \rangle ^{-1}\equiv \tau \sim l^{2}.
\end{equation}
We observe that (11) points out to the signature of the brownian character
of these diffusion limited reactions.

Applying hypothesis (a) to the second term of (6) leads to:

\begin{equation}
\left| -\int_{l^{d}}d^{d}r\left( h\langle \epsilon \rangle \epsilon \right)
\right| \sim \langle h\rangle \langle \epsilon \rangle
^{2} l^{d}\sim 1
\end{equation}
Supposing $\langle h \rangle \sim 1$, we have for the mean
value of $\epsilon $ ($\langle \epsilon $ $\rangle $) the behavior:
\begin{equation}
\langle \epsilon \rangle \sim l^{-\frac{d}{2}}
\end{equation}

Using (11) into (13), we obtain:
\begin{equation}
\langle \epsilon \rangle \sim \left( l^{2}\right) ^{-\frac{d}{4}
}\sim \tau ^{-\frac{d}{4}}
\end{equation}
This result (Eq. (14)) agrees with that obtained by Bramson and Lebowitz
[5] which performed rigorous calculations for the $A+B\rightarrow ~ 0$
reactions. We observe that we have treated here only the case of equal
initial concentrations, namely $\langle \epsilon _{A}\rangle _{0}
=\langle \epsilon _{B}\rangle _{0}$.
The case of unequal initial concentrations will be treated elsewhere.

Now let us use assumption ({\bf a}) in the third term of (6),
\begin{equation}
\left| \int_{l^{d}}d^{d}r\left( \frac{1}{3}K \epsilon ^{3}\right) \right|
\sim \langle K\rangle \langle \epsilon ^{3}\rangle l^{d}\sim 1
\end{equation}
In order to pursue further let us make a plausible hypothesis:
\begin{equation}
\langle \epsilon ^{3}\rangle \sim \langle \epsilon ^{2}\rangle \langle
\epsilon \rangle
\end{equation}
This kind of decoupling could be justified taking in account that the
reaction term which appears in Eq. (5) is essentially of a bilinear form (
a two points correlation). Beside this we observe that this decoupling has
been worked well in a previous calculation [14], where we reproduced the
exact results of Peliti [12] for the $A+A\rightarrow 0$ reaction.

Using (8), (13) and (16) in (15), we obtain:

\begin{equation}
\langle K\rangle \sim l^{\frac{d}{2}-2}\sim \left( l^{2}\right) ^{
\frac{d}{4}-1}
\end{equation}
By considering (11), we can write also:
\begin{equation}
\langle K\rangle \sim \tau ^{\frac{d}{4}-1}
\end{equation}
Equation (17) displays $d=4$, as the upper critical dimension for the model.

We see from ( 17) that $\langle $ $K\rangle $ diverges as $l\rightarrow
\infty $, for $d > 4$. Then, in order to satisfy assumption ({\bf b}) we
adopt $\langle $ $K\rangle $ $=$ $1$, for $d > 4$. So the mean field
description of this model is correct for $d > 4$.

Now, at the upper critical dimension ($d=4$), let us make some additional
considerations in order to obtain the logarithmic corrections to the reactant
concentration $\epsilon $. We have for $d=4$( see (8), (13) and (16)):
\begin{equation}
\langle \epsilon ^{3}\rangle \sim \langle \epsilon ^{2}\rangle \langle
\epsilon \rangle \sim l^{-4}, ~~~for~~~ d=4.
\end{equation}

If we use assumption ({\bf a}) for the third term of (6) in a somewhat modified
context, and after having made inside the integral the following
substitution:
\begin{equation}
\epsilon ^{3}\sim r^{-4}
\end{equation}
we can write
\begin{equation}
\int_{l^{d}}d^{d}r\left( \frac{1}{3}K\epsilon ^{3}\right) \sim
\int_{1}^{l}\left( \frac{1}{3}Kr^{-4}\right) r^{3}dr\sim \langle
K\rangle \ln \left( l\right) \sim 1
\end{equation}
where $l$ is measured in units of space lattice, being $1$ the lower cutoff.

Then we have for $d=4$,
\begin{equation}
\langle K\rangle \sim \left[ \ln \left( l\right) \right] ^{-1}.
\end{equation}
At this point, let us consider the following differential equation:
\begin{equation}
\frac{1}{2}D\left( \frac{\partial \epsilon }{\partial r}\right) ^{2}-\frac{1
}{3}\langle K\rangle \epsilon ^{3}=0.
\end{equation}
This equation can be obtained by considering the equality of the first and
third terms of (6)( given by assumption a), where we suppose that we can
substitute equality between integrals by equality between integrands, after
replacing $K$ by $\langle K\rangle $.

We can solve ( 23) at the upper critical dimension ( $d=4$), by performing
the integration between $1$ and $l$, and using (22). We get:
\begin{equation}
\langle \epsilon \rangle \sim \frac{\ln \left( l\right) }{l^{2}}
\sim \frac{\ln \left( \tau \right) }{\tau }, ~~ {\rm for} ~~ d=4.
\end{equation}
As we can see from (22) just at the upper critical dimension for the
$A+B\rightarrow $\ $0$ reaction ($d_{c}=4$), we have obtained a logarithmic
correction to the mean field description. The same behavior was obtained
before for the $A+A\rightarrow 0$ reaction at $d_{c}=2$.(see [14]).

\bigskip

\section
{Unified Treatment of $A+A \rightarrow 0$ and 
$A+B\rightarrow 0$ reactions.}

The relative success of Thompson's approach [15] in treating both
$ A + A \rightarrow ~ 0$ and $A + B \rightarrow ~ $0 reactions
encourage us to look for an
extended action which could encompass the two mentioned reactions as
limiting cases. So let us write the following extended action.
\begin{equation}
A_{\sigma }=\int_{l^{d}}d^{d}r\left[ \frac{1}{2}D\left( \nabla \epsilon
\right) ^{2}-h\langle \epsilon \rangle ^{\sigma }\epsilon +\frac{1}{3}
K\epsilon ^{3}+\frac{1}{2}\frac{\partial \left( \epsilon ^{2}\right) }{
\partial t}\right].
\end{equation}
where $\sigma \ge 0$.

In an analogous ways that we have worked before, we can apply Thompson's
method [15] to the action giving by (25), obtaining:
\begin{equation}
\langle \epsilon \rangle _{\sigma }\sim l^{-d /(\sigma +1)},
\end{equation}
and
\begin{equation}
\langle K\rangle _{\sigma }\sim \left( l^{2}\right) ^{\frac{d}{2\sigma
+2}-1}.
\end{equation}
Relation (27) implies that, for the model defined in (25), the upper
critical dimension is given by
\begin{equation}
d_{c}\left( \sigma \right) =2\sigma +2.
\end{equation}
So alternatively we can write $\langle K\rangle _{\sigma }$ as
\begin{equation}
\langle K\rangle _{\sigma }\sim \left( l^{2}\right) ^{\frac{d}{d_{c}}
-1}\sim \tau ^{\frac{d}{d_{c}}-1}.
\end{equation}
We observe that if $\sigma =0$, $d_{c}=2$ and we recover the
$A+A\rightarrow ~ 0$ behavior [12,14], and putting $\sigma =1$ we get
$d_c= 4$, reproducing the results of the previous section for the
$A+B\rightarrow 0$ reactions.

It is interesting to verify that the action given by (25) can reproduce the
results of the LSK work [17], in the case of euclidian geometries. First we
can write:
\begin{equation}
\langle \epsilon \rangle _{\sigma }\sim \tau^{-d/(2\sigma +2)}
=\tau^{-d/d_{c}}.
\end{equation}
Making the requirement that the exponent of right side of (30) to be equal
to $-\alpha $ given by (4), and by setting $d_{s}=d_{f}=d$, we have:
\begin{equation}
\sigma =\frac{\gamma }{2-\gamma }.
\end{equation}
Therefore putting $\gamma =\sigma =0$, and $\gamma =\sigma =1$, we get the
long time behavior of the $A+A\rightarrow ~ 0$ and
$A+B\rightarrow ~ 0$ reactions respectively. The case of $A+B\longrightarrow
~ 0$ reactions where correlations are important (see [17]) can be
described for some intermediate values of the parameters $\gamma $
(or $\sigma $).

Finally is worth to point out that at the upper critical dimension of the
model described by the action (25) we have again the
logarithmic corrections to the mean field behavior. This result can be
obtained in the same way we have worked out before in section 2. Doing this,
we obtain:
\begin{equation}
\langle \epsilon \rangle _{\sigma }\sim \frac{\ln \left( l\right) }{
l^{2}}\sim \frac{\ln \left( \tau \right) }{\tau }, ~~ {\rm for} ~~ 
d_{c}=2\sigma +2.
\end{equation}

Before concluding this section we would like to write an extendend version of
the renormalized reaction rate in the same spirit of Peliti's work [12]. For
long time and long wavelengths, we can write the scaling relation:
\begin{equation}
K_{\sigma }\sim \left( l^{2}\right) ^{\frac{d}{2\sigma +2}-1}.f\left( 
\frac{l^{2}}{\tau }\right).
\end{equation}
Putting $\sigma =0$, we recover the renormalized reaction rate of Peliti
[12] ( equation (10) of his paper).

\section{Concluding remarks.}

In this paper we have proposed a model lagrangian ( action) as a means to
describe a large class of diffusion limited reactions. The long time
behavior of the concentration $\epsilon $ and of the reaction rate $K$ were
evaluated by using Thompson's approach [15],which can be thought as being a
simple alternative way to the renormalization group method [16].

In two particular cases of this extended action, we recover the exact
results of Peliti [12] for $A+A\longrightarrow 0$ and the rigorous bounds of
Bramson and Lebowitz[5] for the $A+B\rightarrow 0$ reactions. Besides its
simplicity, one of the advantages of the present method is that it can
encompass diffusion limited reactions of different classes of universality
within the same formalism. Moreover, our calculations display logarithmic
behavior of the species concentrations and reactions rates at their
respectives upper critical dimensions for any class of universality.

An other advantage of the present method is that it can be extended to
study, for example, the scalar field theories $\Phi ^{3}$, $\Phi ^{4}$,
and $\Phi ^{n}$ for dimension d, in such a way that we can obtain a certain
critical dimension $d_{c}$ for a given value of $n$ in $\Phi ^{n}$ theory in a
closed form. But now, $\ d_{c}$ must be interpreted as a dimension
where the $\Phi ^{n}$ theory becomes renormalizable displaying logarithmic
dependence of the coupling constant in the energy scale. This line of
reasoning will be subject of a forthcomming paper.

\end{document}